\documentclass[]{piparticle-final}
\usepackage{graphicx}
\usepackage{amsmath}

\begin{document}
\volume{2}               
\articlenumber{020005}   
\journalyear{2010}       
\editor{D. H. Zanette}   
\reviewers{V. M. Eguiluz, Inst. F\'{\i}sica Interdisciplinar y Sist. Complejos, \\ \mbox{}\hspace{42.5mm}Palma de Mallorca, Spain}  
\received{17 July 2010}     
\accepted{27 September 2010}   
\runningauthor{ J. I. Perotti \itshape{et al.}}  
\doi{020005}         

\title{Stability as a natural selection mechanism on interacting networks}

\author{Juan I. Perotti,\cite{inst1,inst2}\thanks{E-mail: juanpool@gmail.com}\hspace{2mm}
        Orlando V. Billoni,\cite{inst1,inst2}\thanks{E-mail: billoni@famaf.unc.edu.ar}\hspace{2mm}
        Francisco A. Tamarit,\cite{inst1,inst2}\thanks{E-mail: tamarit@famaf.unc.edu.ar}\hspace{2mm}
        \\Sergio A. Cannas\cite{inst1,inst2}\thanks{E-mail: cannas@famaf.unc.edu.ar}}

\pipabstract{Biological networks of interacting agents exhibit similar topological properties for a wide range of scales, from cellular to ecological levels, suggesting the existence of a common evolutionary origin. A general evolutionary mechanism based on global stability has been proposed recently  [J I Perotti, {\it et al.}, Phys. Rev. Lett. {\bf 103}, 108701 (2009)]. This mechanism was incorporated into a  model of a growing network of interacting agents in which each new agent's membership in the network is determined by the agent's effect on the network's global stability. In this work, we analyze different quantities that characterize the topology of the emerging networks, such as global connectivity, clustering and average nearest neighbors degree, showing that they reproduce scaling behaviors frequently observed in several biological systems. The influence of the stability selection  mechanism on the dynamics associated to the resulting network, as well as  the interplay  between some topological and functional features are also analyzed.}


\maketitle

\blfootnote{
\begin{theaffiliation}{9}
   \institution{inst1} Facultad de Matem\'atica, Astronom\'{\i}a y F\'{\i}sica,
Universidad Nacional de C\'ordoba, Argentina.
\institution{inst2}  Instituto de F\'{\i}sica Enrique
Gaviola  (IFEG-CONICET),
Ciudad Universitaria, 5000 C\'ordoba, Argentina.
\end{theaffiliation}
}

\section{Introduction}

The concept of  networks of interacting agents has proven, in the last decade, to be a powerful tool in the analysis of complex systems (for reviews, see Refs. [1-4]). 
Although not new, with the advent of high performance computing, this theoretical construction opened a new door for the statistical physics methodology in the analysis of systems composed by a large number of units that interact in a complicated way. This allowed to get new insights about the dynamical behavior of systems as complex as biological and social systems. In addition, it constitutes a basic backbone  upon which relatively simple models can be constructed in a  bottom-up  strategy.

As a modeling tool, the definition of an interaction network for a given system is frequently not unique (see for example the case of protein-protein interaction networks [5-7]), 
 depending on the coarse grain level of the approach. Nevertheless, many topological properties appear to be independent of the definition of the network. Moreover, some of those  properties have emerged in the last years as universal features among systems otherwise considered very different from each other. In particular, the following properties are characteristic of most biological networks. \textit{(a) Small worldness}:  all of them exhibit high
clustering $Cc$ and relatively short path length $L$, compared with random networks.
$L$ is defined as the minimum number of links needed to connect
any pair of nodes in the network and $Cc$ is defined as the
fraction of connections between topological neighbors of any
site\cite{AlBa2002}. \textit{(b) Scale free degree distribution}: the degree distribution $P(k)$ (the probability of a node to be connected to $k$ other ones) presents a broad tail   for large values of $k$. In some cases, the tail can be approached by a power law $P(k) \sim k^{-\gamma}$ with
degree exponents $\gamma<3$ for a wide range of scales, while in others, a cutoff appears for some maximum degree $k_{max}$; in the latter, the degree distribution is generally well described by  $P(k) \sim k^{-\gamma}\, e^{-k/k_{max}}$ [1, 7-12]. 
In any case, the networks present a nonhomogeneous structure, very different from that expected in a random network. \textit{(c) Scaling of the clustering coefficient}: in many natural networks, it is observed that the clustering coefficient of a node with degree $k$ follows the scaling law $Cc(k) \sim k^{-\beta}$, with $\beta$ taking values close to one. This has been interpreted as an evidence for a modular structure organized in a hierarchical way \cite{RaBa2003}. \textit{(d) Disassortative mixing by degree}: in most biological networks, highly connected nodes tend to be preferentially connected to nodes with low degree and vice versa \cite{Ne2003}.

These properties are observed for a wide
range of scales, from the microscopic level of genetic, metabolic
and proteins networks to the macroscopic level of communities of living beings (ecological networks). Such ubiquity suggests the existence  of some natural selection  process that promotes the development of those particular structures \cite{PrPrPh2005}.  One possible constraint general enough to act across such a range of scales is the  proper stability of the underlying dynamics.

Growing biological networks involve the
coupling of at least two dynamical processes. The first one
concerns the addition of new nodes, attached during a slow
evolutionary (i.e., species lifetime) process. A second one is
the node dynamics which affects and in turn is affected by the
growing processes. It is reasonable to expect that the
network topologies we finally witness could have emerged
out of these coupled processes. Consider, for example, the case of an ecological network like a
food web, where nodes are species within an ecosystem and edges are consumer-resource relationships between them. New nodes are added during evolutionary time scales, through speciation or migration of new species. Then, the network grows through community assembly rules, strongly
influenced by the underlying dynamics of species and specific
interactions among them \cite{WeKe1999,Pi1991}.
 The consequence of adding a new member
with a given connectivity affecting a global in/stability, is
represented in this case by the aboundance/lack of food \cite{Note-food-webs}. Notice that each new member may not only result in its own
addition/rejection to the system, but it can also promote avalanches
of extinctions amongst existing members.

The above ingredients were recently incorporated into a simple
 model of growing networks under stability constraints \cite{PeBiTaChCa2009}. Numerical simulations on this model  showed that, indeed, complex topology can emerge out of a stability selection pressure. In the present work, we further explore different topological and dynamical properties predicted by the model, whose definition is reviewed  in section \ref{model} The results are presented in sections \ref{topology} and \ref{structure} In section \ref{topology}, we analyze the topological features that emerge in growing networks under stability constraint. In section \ref{structure}, we show that this constraint not only induces  topological features of the resulting networks but also influences the associated dynamics.  A discussion of the results is presented in section \ref{conclu}

\section{The Model}
\label{model}

Let us consider a system of $n$ interactive agents, whose
dynamics is given by a set of differential equations $d \vec{x}/dt= \vec{F}(\vec{x})$, where $\vec{x}$ is an $n$-component vector describing
the relevant state variables of each agent and $\vec{F}$ is an
arbitrary non-linear function.  One could imagine that $\vec{x}$ in
different systems  may represent concentrations of some hormones,
the average density populations in a food web, the
concentration of chemicals in a biochemical network, or the
activity of  genes in a gene regulation net, etc. We  assume that
a given agent $i$ interacts only with a limited set of $k_i < n$
other agents; thus, $F_i$ depends only on the variables belonging to that
set. This defines the interaction network.

We  assume that there are two time scales in the dynamics. Let $f_m$ be the average frequency of the incoming flux of new agents (migration, mutation,
etc.). This defines a characteristic time $\tau_m = f_m^{-1}$.  On
the long time scale $t\gg \tau_m$ (much larger than the observation time) new agents arrive to the system and start to interact with some of the previous ones. Some of them
 can be incorporated into the system or not, so $n$ (and the
whole set of differential equations) can change. Once a new agent starts to interact with the system, we will assume that the enlarged system evolves towards some stationary state with characteristic relaxation time $\tau_{rel} \ll \tau_{m}$. Then, in the short time
scale $\tau_{rel} \ll t\ll \tau_m$ we can assume that $n$ is constant and the dynamics already
led the system to a particular stable stationary state
$\vec{x}^*$ defined by $\vec{F}(\vec{x}^*)=0$. Following May's ideas \cite{Ma1972}, we assume that the only attractors of the dynamics are fixed points. Nevertheless, the proposed mechanism is expected to work, as well,
for more complex attractors (e.g, limit cycles).

The stability of
the solution $\vec{x}^*$ is determined by the eigenvalue with maximum real
part of the Jacobian matrix

\begin{align}
a_{i,j}\equiv \left(\frac{\partial
F_i}{\partial x_j}\right)_{\vec{x}^*}\,.
\end{align}

\noindent A new agent will be
incorporated to the network if its inclusion results in a new stable
fixed point, that is, if the values of the interaction matrix
$a_{i,j}$ are such that the eigenvalue with maximum real part
$\lambda_m$ of the enlarged Jacobian matrix is negative ($\lambda_m<0$). Assuming
that isolated agents will reach  stable states by themselves after
certain characteristic relaxation time, the diagonal elements of
the matrix ${a_{i,i}}$ are negative and given unity value to
further simplify the treatment \cite{Ma1972}. The interaction values, (i.e.,
the non-diagonal matrix elements ${a_{i,j}}$) will take random
values (both positive and negative) taken from some statistical
distribution.  In this way, we have an unbounded ensemble of systems \cite{Ma1972} characterized by a ``growing through stability'' history. Randomness  would be self-generated through the addition of new agents processes. Each specific set of matrix elements, after addition, defines a particular dynamical system and the subsequent analysis for time scales between successive migrations is purely deterministic.

The model is then defined by the following algorithm \cite{PeBiTaChCa2009}. At
every step, the network can either grow or shrink. In each step, an
attempt is made to add a new node to the existing network,
starting from a single agent ($n=1$).  Based on the stability
criteria already discussed, the attempt can be successful or not. If
successful, the agent is accepted, so the existing $n \times n$
matrix grows its size by one column and one row. Otherwise, the
novate agent will have a probability to be deleted together with
some other nodes, as further explained below.

 More specifically, suppose that we have an already created network
with $n$ nodes, such that the $n \times n$ associated interaction
matrix ${a_{i,j}}$ is stable. Then, for the attachment of the
$(n+1)_{th}$ node, we first choose its degree $k_{n+1}$ randomly between $1$
and $n$ with equal  probability. Then, the new agent interaction
with the existing network member $i$ is chosen, such that non-diagonal
matrix elements $(a_{i,n+1},a_{n+1,i})$ ($i=1,\ldots,n$) are zero
with probability $1-k_{n+1}/n$ and different from zero with probability
$k_{n+1}/n$; to each non--zero matrix element we assign a different real random value uniformly
distributed in $[-b,b]$. $b$ determines the interaction range
variability and it is one of the two parameters of the model \cite{Note1}.

Then, we calculate numerically $\lambda_m$ for the resulting $(n+1)
\times (n+1)$ matrix. If $\lambda_m<0$,   the
new node is accepted. If $\lambda_m>0$,
it means that the introduction of the new node destabilized the
entire system and we will impose that, the new agent is either
eliminated or it remains but produces the extinction of a certain
number of previous existing agents.
In order to further simplify the numerical treatment, we
allow up to $q \leq k_{n+1}$ extinctions, taken from the set of
$k_{n+1}$ nodes connected to the new one; $q$ is
the other parameter of the model. To choose which nodes are to be
eliminated, we first select one with equal probability in the set of
$k_{n+1}$ and remove it. If the resulting $n \times n$ matrix is
stable, we start a new trial; otherwise, another node among the
remaining $k_{n+1}-1$ is chosen and removed, repeating the
previous procedure. If after $q$ removals the matrix remains
unstable, the new node is removed (we return to the original
$n\times n$ matrix and start a new trial). The process is repeated until the network reaches a maximum size $n=n_{max}$ (typically $n_{max}=200$) and restarted $M$ times from $n=1$ to obtain statistics of the networks (typically $M=10^5$).

\section{Topological properties}
\label{topology}

\subsection{Connectivity}

\begin{figure}
\begin{center}
\includegraphics[scale=0.28,angle=-90]{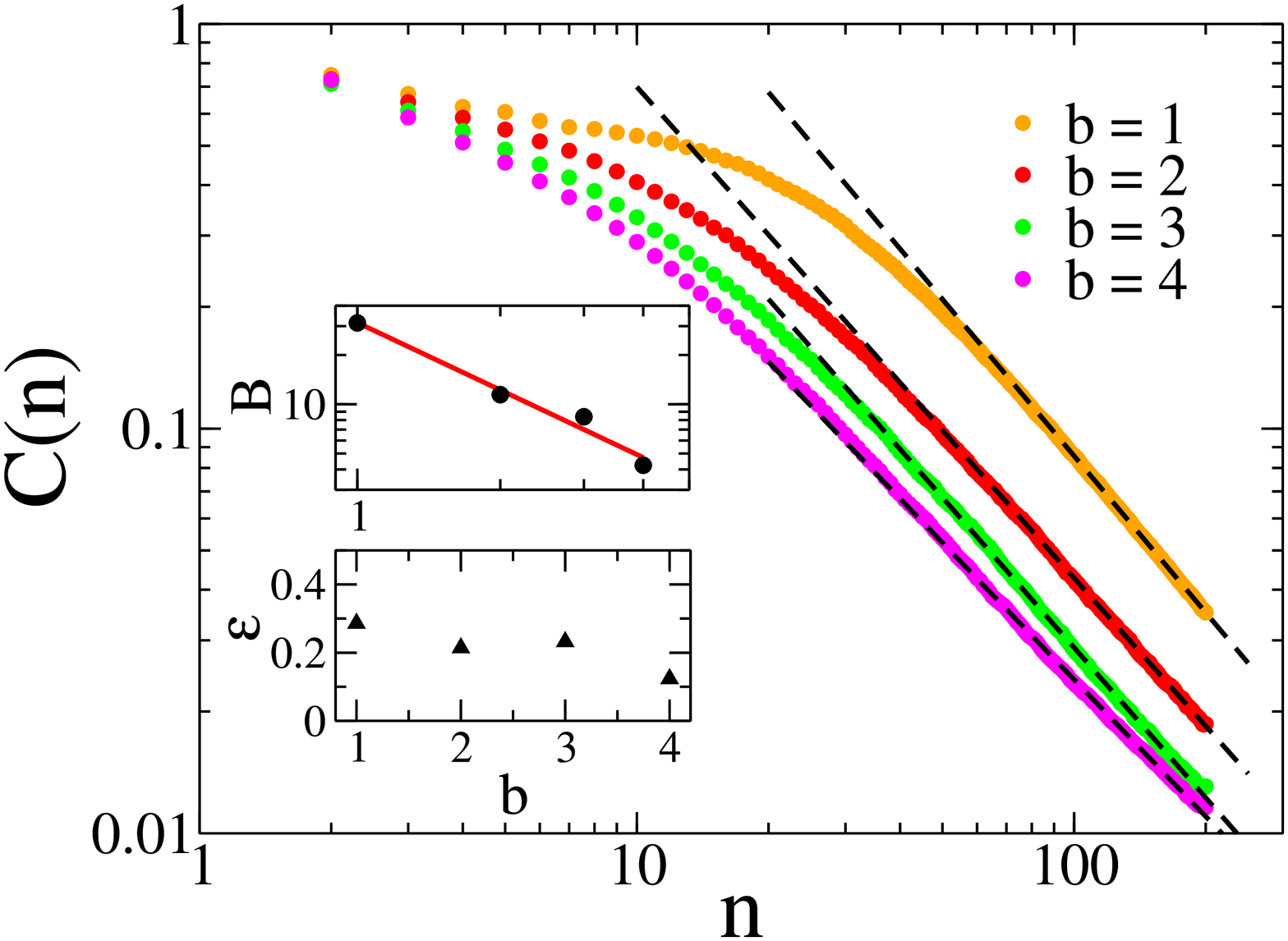}
\caption{\label{fig1} Connectivity as a function of the network size for $q=3$, $n_{max}=200$ and different values of $b$. The symbols correspond to numerical simulations and the dashed lines to power law fittings of the tails $C(n) = B\, n^{-(1+\epsilon)}$. The insets show the fitting values $B$ and $\epsilon$ as a function of $b$}
\end{center}
\end{figure}

First, we analyzed the average connectivity $C(n)$, defined as
the fraction of non-diagonal matrix elements different from zero,
averaged over different runs.  In Fig.~\ref{fig1}, we show the typical behavior of $C(n)$ for different values of $b$ (we found that $C(n)$ is completely independent of $q$). The connectivity presents a power law tail for large values of $n$. From a fitting of the tail with a power law (see insets in Fig.~\ref{fig1}) we obtain the scaling behavior

\begin{align}\label{connectivity}
    C(n)\sim
\alpha^{-\omega}\,n^{-(1+\epsilon)}
\end{align}

\noindent  for large values of
$n$, where $\alpha$  is the variance of the non-diagonal elements of the stability matrix  ($\alpha=b^2/3$ for the uniform distribution) and $\omega = 0.7 \pm 0.1$. From the inset of Fig.~\ref{fig1}, we see that the exponent $\epsilon$ shows a weak dependency on $b$, taking values in the range $(0.1,0.3)$ . It is interesting to compare Eq.~(\ref{connectivity}) with May's stability line for random networks \cite{Ma1972} $C(n)~=~(\alpha  n)^{-1}$. It is easy to see that Eq.~(\ref{connectivity}) lies above May's stability line for network sizes up to $\sim 10^6$ \cite{Note2}. This shows that networks growing under stability constraint develop particular structures whose probability in a completely random ensemble is almost zero. In other words, the associated matrices belong to a subset of the random ensemble with zero measure and therefore they are only attainable through a constrained development process. In the next sections, we  explore the characteristics of those networks.

\begin{figure}
\begin{center}
\includegraphics[scale=0.28,angle=-90]{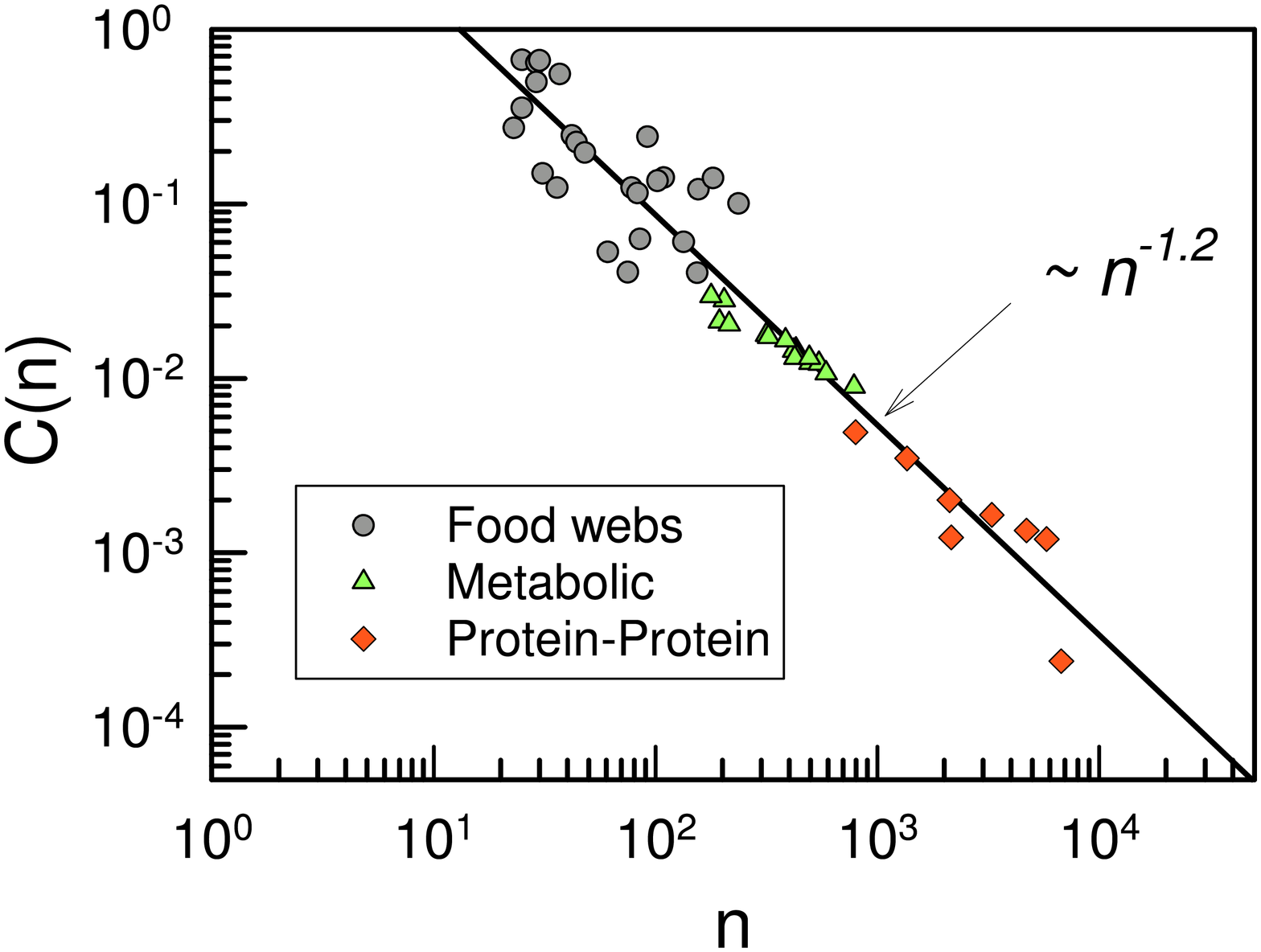}
\caption{\label{Cndata}  Connectivity as a function of the network size for different biological networks. The straight line is a power law fitting $C(n) = A\, n^{-(1+\epsilon)}$, giving an exponent $\epsilon = 0.2 \pm 0.1$ ($R^2=0.92$). Data extracted from:  \cite{YoOlBa2004,CoFlMaVe2005} (protein-protein interaction networks); \cite{JeToAlOlBa2000} (metabolic networks); \cite{MeBa2004,Du2009} (food webs). }
\end{center}
\end{figure}

In Fig.~\ref{Cndata}, we plotted the connectivity  for different biological networks across three orders of magnitude of network size scales, using data collected from the literature. We see that the data are very well fitted by a single power law $C(n) \sim n^{-1.2}$, in a nice agreement with the average value $\epsilon=0.2$ predicted by the present model. It is worth mentioning that the behavior $C(n) \sim n^{-(1+\epsilon)}$ has also been obtained in a self organized criticality model of Food Webs \cite{SoAlMc2000}.

\subsection{Degree distribution}
\label{degree}
\begin{figure}
\begin{center}
\includegraphics[scale=0.28,angle=-90]{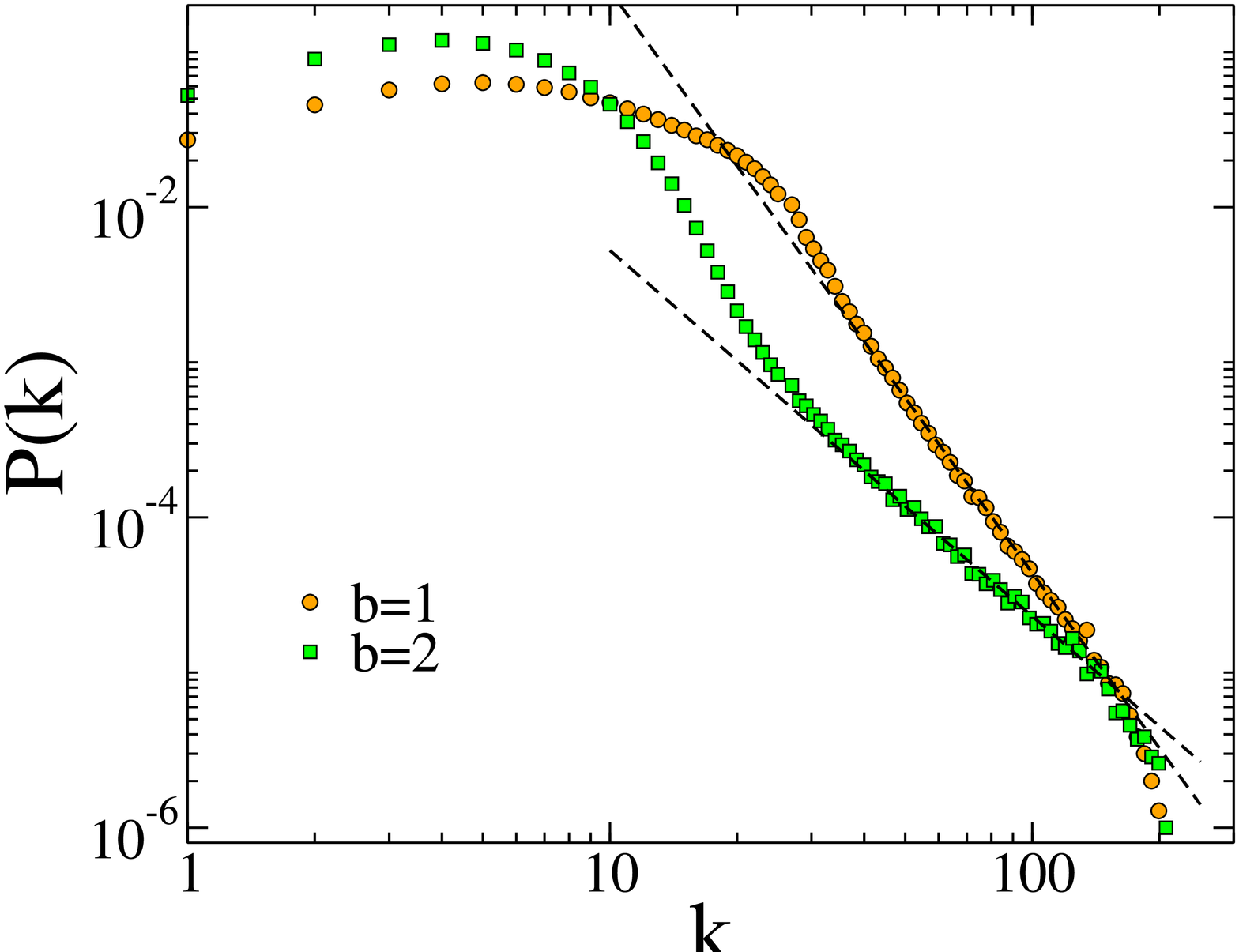}
\caption{\label{fig2} Degree distribution $P(k)$
for $q=3$, $n_{max}=200$ and different values of $b$; the dashed
lines correspond to power law fittings of the tail $P(k) \sim k^{-\gamma}$. Logarithmic binning has been used to smooth the curves.}
\end{center}
\end{figure}

The degree distribution $P(k)$ of the network was analyzed in detail in Ref. \cite{PeBiTaChCa2009}. We briefly summarize the main results here. In Fig.~\ref{fig2}, we illustrate the typical behavior of $P(k)$.
 It presents  a power law tail $P(k) \sim k^{-\gamma}$ for values of $k > 20$, with a finite size drop at $k=n_{max}$. The degree exponent $\gamma$  takes values between $2$ and $3$ for values of $b$ in the interval $b \in (1.5,3.5)$, which become almost independent of $q$ as it increases. The exponent $\gamma$ can also fall below $2$ when the global stability constraint is replaced by a local one.  The qualitative structure of $P(k)$ remains when the stability criterium $\lambda_m < 0$ is relaxed by the condition $\lambda_m < \Delta$, with $\Delta$ some small positive number. In other words, the power law tail emerges also when the addition of new nodes destabilizes the dynamics, provided that the characteristic time to leave the fixed point $\tau=\lambda_m^{-1}$ is large enough to become comparable to the migration time scale $\tau_m$ \cite{PeBiTaChCa2009}.

\subsection{Network growth and clustering properties}

\begin{figure}
\begin{center}
\includegraphics[scale=0.28,angle=-90]{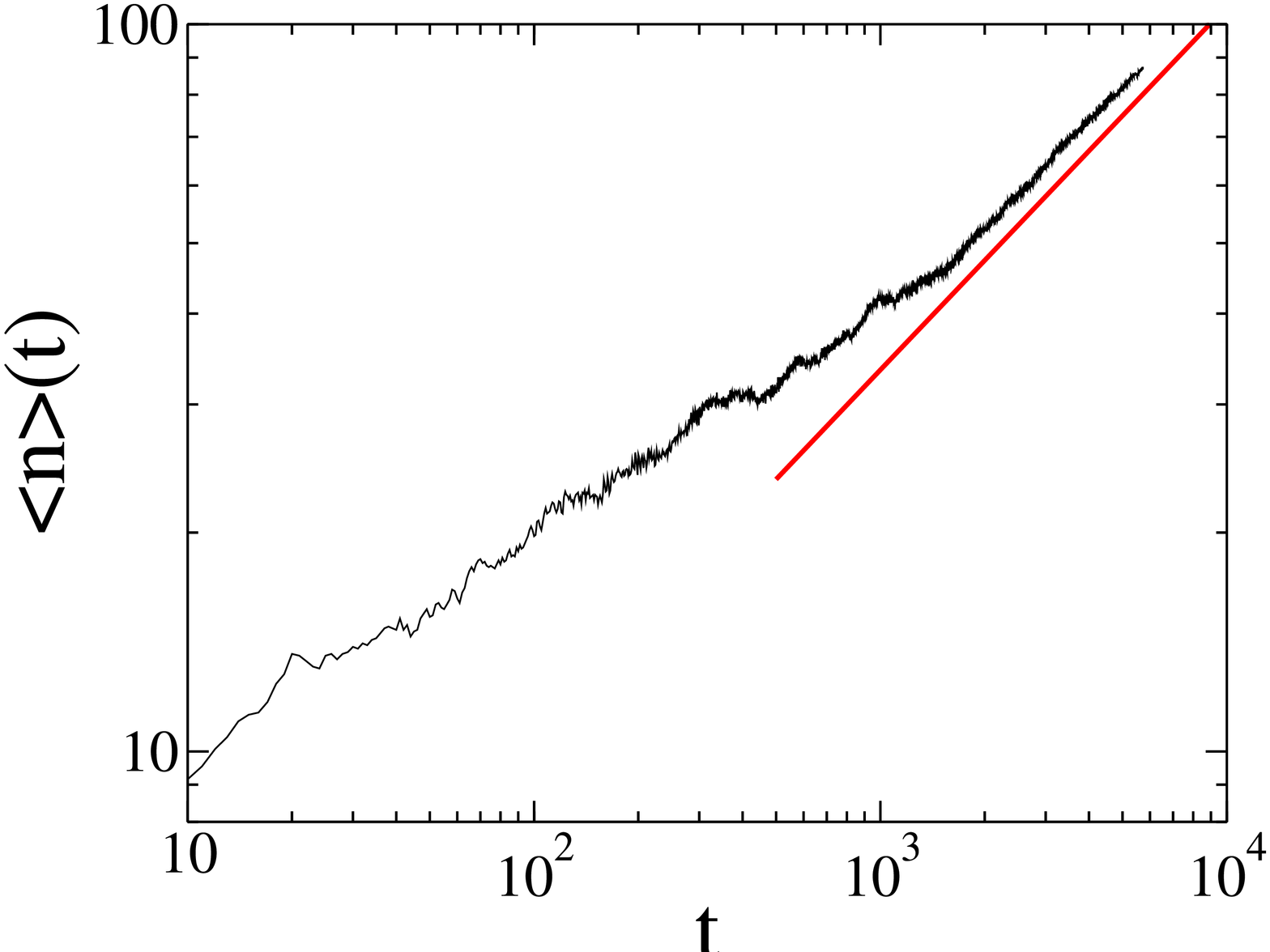}
\caption{\label{growth}  Average network size as a function of the time measured in number of trials for $b=2$ and $q=3$. The continuous red line corresponds to the power law $t^{1/2}$,}
\end{center}
\end{figure}

Networks grown under stability constraint also display small world properties. The average clustering coefficient decays with the network size as $Cc(n) \sim
n^{-0.75}$ (which is slower than the $1/n$ decay in a random net), while the average path length $L$ between two nodes increases as $L(n) \sim A\, \ln{(n+C)} $ \cite{PeBiTaChCa2009}. A similar behavior is observed in the  Barabasi-Albert model \cite{AlBa2002}, where the clustering can be approximated by a power law with the same exponent, although the exact scaling is \cite{KlEg2002} $Cc(n) \sim (\ln n)^2/n$ (therefore that behavior cannot be excluded in the present model). While this suggests the presence of an underlying preferential attachment rule  mechanism, a detailed analysis has shown that this is not the dominant mechanism \cite{PeBiTaChCa2009}.
 The
behavior of $Cc$ and $L$ is linked with the selection dynamics
ruling which node is accepted or rejected. The stability
constraint favors the nodes with few links, since they modify the
matrix ${a_{i,j}}$ stability much less than new nodes with many
links (of course this is reflected in the $P(k)$ density). Thus,
most frequently, the network grows at the expense of adding nodes
with one or few links, producing an increase of $L$ and a
decrease of $Cc$, but sporadically,  a highly connected node is accepted, decreasing $L$ and increasing $Cc(n)$ \cite{PeBiTaChCa2009}. Those fluctuations lead to a slow
diffusive-like growth of the network size $n(t)\sim t^{1/2}$ (See Fig.~\ref{growth}).

\begin{figure}
\begin{center}
\includegraphics[scale=0.28,angle=-90]{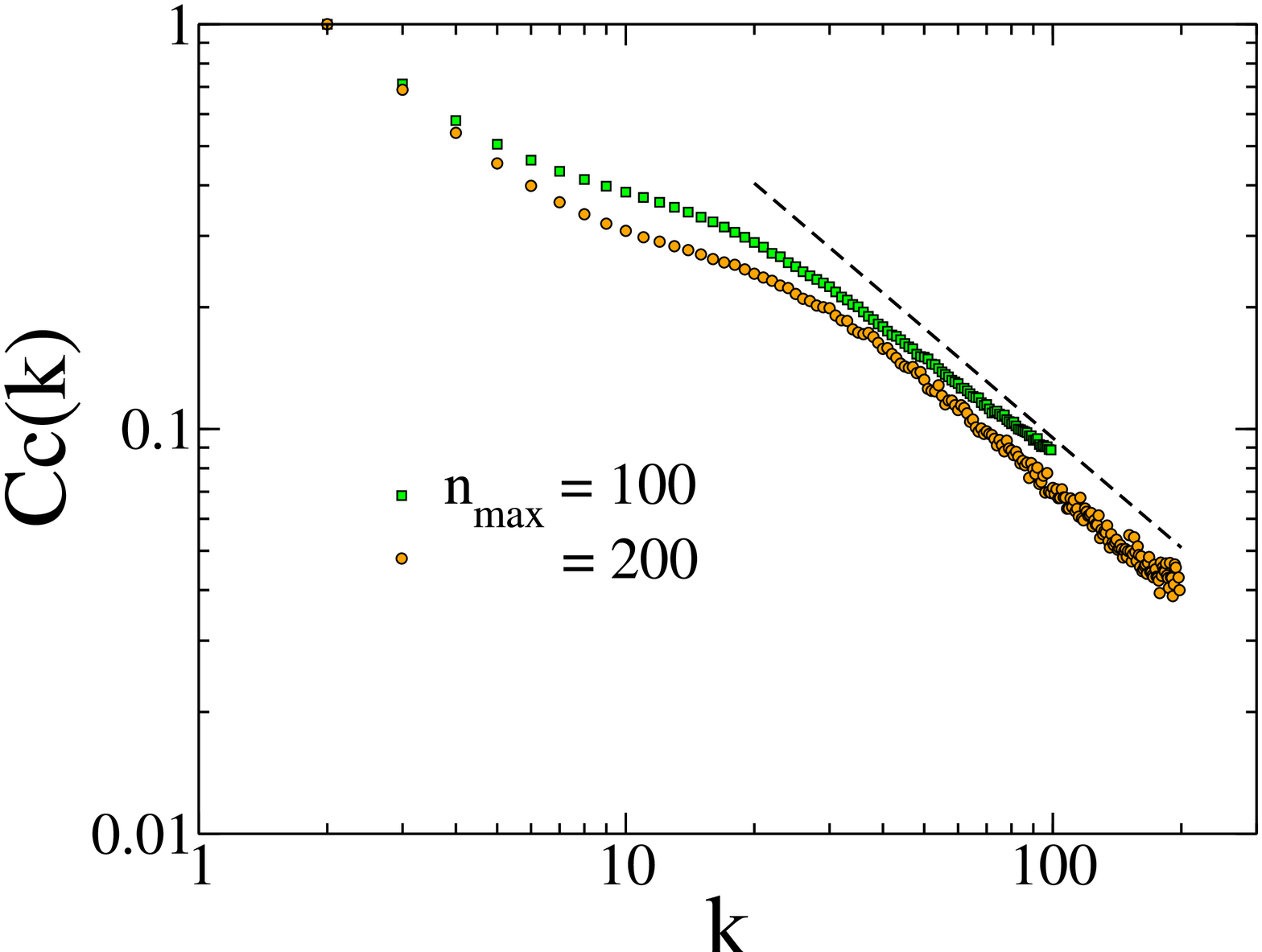}
\caption{\label{Cck}  Average clustering coefficient $Cc(k)$ as a function of the degree $k$ for $b=2$, $q=3$ and different values of $n_{max}$. The dashed black line corresponds to the power law $k^{-0.9}$.}
\end{center}
\end{figure}

 Another quantity of interest is the average clustering $Cc(k)$ as a function of the  degree $k$. A typical example is shown in Fig.~\ref{Cck}. We see that $Cc(k)$ decreases monotonously with $k$ and displays a power law tail $Cc(k) \sim k^{-\beta}$ with an exponent $\beta\approx 0.9$, close to one. The exponent appears to be completely independent of $b$ and $q$. This behavior is indicative of a modular structure with hierarchical organization \cite{RaBa2003}.  Notice that this power law decay appears for degrees $k>20$, precisely the same range of values for which the degree distribution $P(k)$ displays  a power law tail (see subsection \ref{degree}).

\subsection{Mixing by degree patterns}

\begin{figure}
\begin{center}
\includegraphics[scale=0.28,angle=-90]{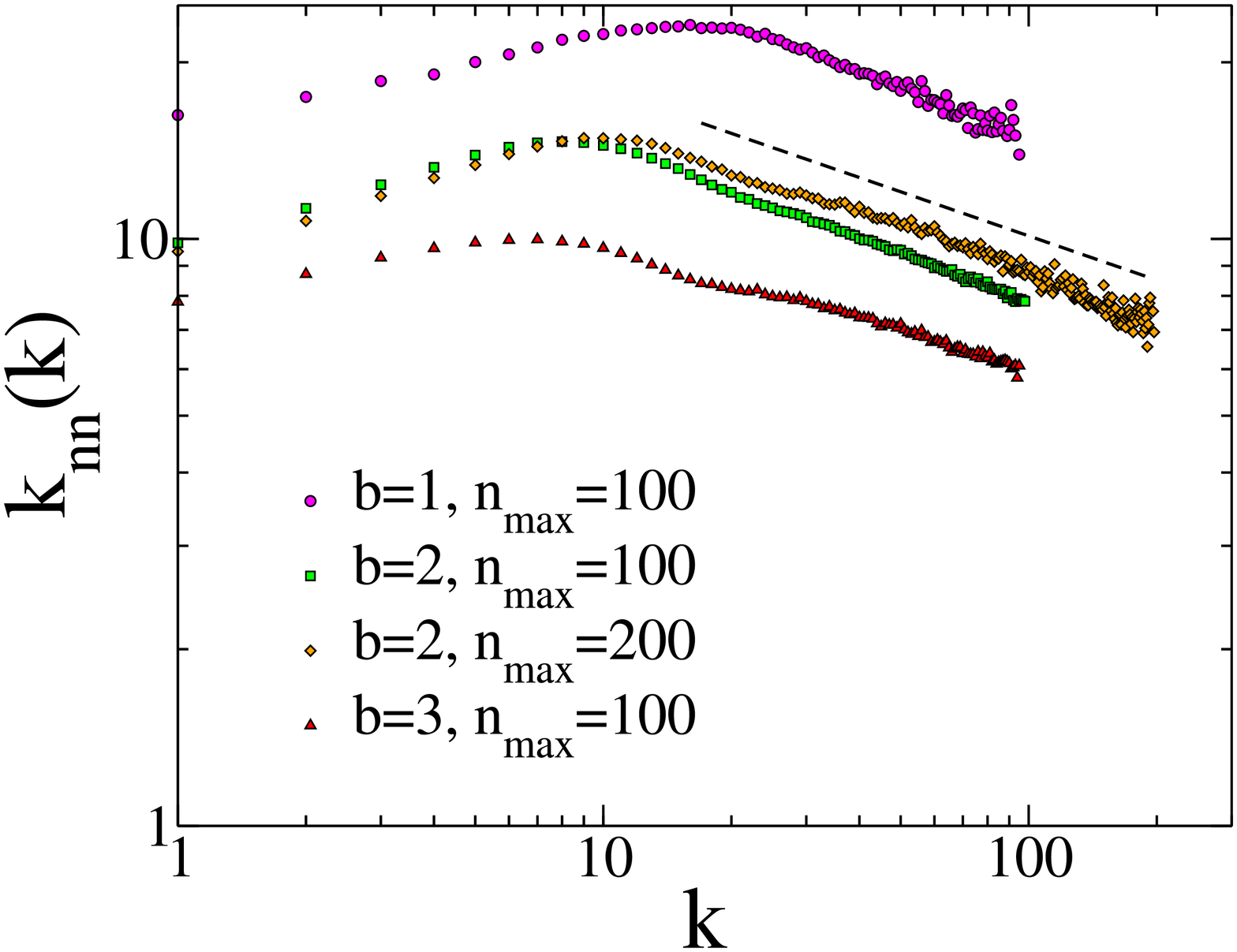}
\caption{\label{Knn}  Average nearest neighbors degree  $k_{nn}(k)$ of a node with degree $k$ for  $q=3$ and different values of $b$ and $n_{max}$. The dashed line corresponds to the power law $k^{-0.25}$.}
\end{center}
\end{figure}

To analyze the mixing by degree properties of the networks selected by the stability constraint, we calculated the
average degree $k_{nn}$ among the nearest neighbors  of a node with  degree $k$. In Fig.~\ref{Knn}, we  see that $k_{nn}$ decays with a power law $k_{nn}\sim k^{-\delta}$ for $k>20$, with an exponent $\delta$ close to $-0.25$, in a clear disassortative behavior.  This result is also consistent with previous works showing that assortative mixing by degree decreases the stability of a network, i.e., the maximum real part $\lambda_m$ of the eigenvalues of random matrices of the type here considered increases faster on assortative networks than on disassortative ones \cite{BrSi2005}.

\section{Dynamical properties}
\label{structure}

In the previous section, we analyzed different topological properties that are selected by the stability constraint, i.e., properties associated to the underlying adjacency matrix, regardless of the values of the interaction strengths. We now analyze the characteristics of the  dynamics  associated to the networks  emerging from such constraint. In other words, we investigate the statistics of values of the non null elements  $a_{ij} \neq 0$.

\begin{figure}
\begin{center}
\includegraphics[scale=0.27,angle=-90]{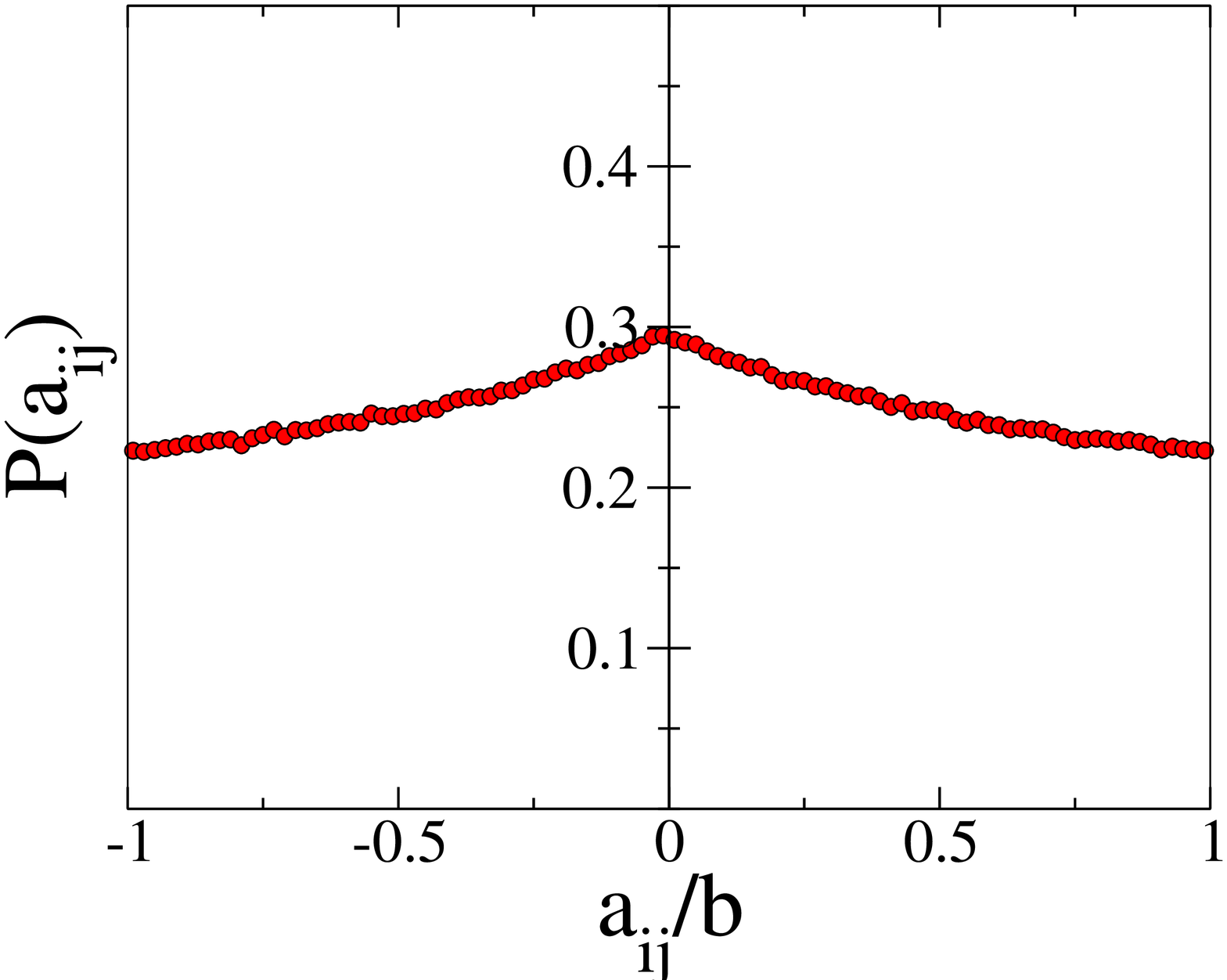}
\caption{\label{aij}  Probability density of the matrix elements $a_{ij}$ for $b=2$, $q=3$ and  $n_{max}=100$. }
\end{center}
\end{figure}

First of all, we calculate the probability distribution of values for a single non null matrix element $a_{ij}$ of the final network with size $n=n_{max}$. The typical behavior is shown in Fig.~\ref{aij}.  We see that $P(a_{ij})$ is an even function, almost uniform in the interval $[-b,b]$, with a small cusp around $a_{ij}=0$. This shows that stability is not enhanced by a particular sign or absolute value  of the individual interaction coefficients. It has been shown recently that the presence of anticorrelated links between pairs of nodes (i.e., links between pairs of nodes $(i,j)$ such that $sign(a_{ij})=-sign(a_{ji}$)) significatively enhances the stability of random matrices \cite{AlPa2008}. In an ecological network, this typically corresponds to a predator-prey or parasite-host interaction. To check for the presence of such type of interactions, we calculated the correlation  $\langle a_{ij}a_{ji} \rangle$, where the average is taken over pairs of nodes with a double link ($a_{ij}\neq 0$ and $a_{ji}\neq 0$).

\begin{figure}
\begin{center}
\includegraphics[scale=0.27,angle=-90]{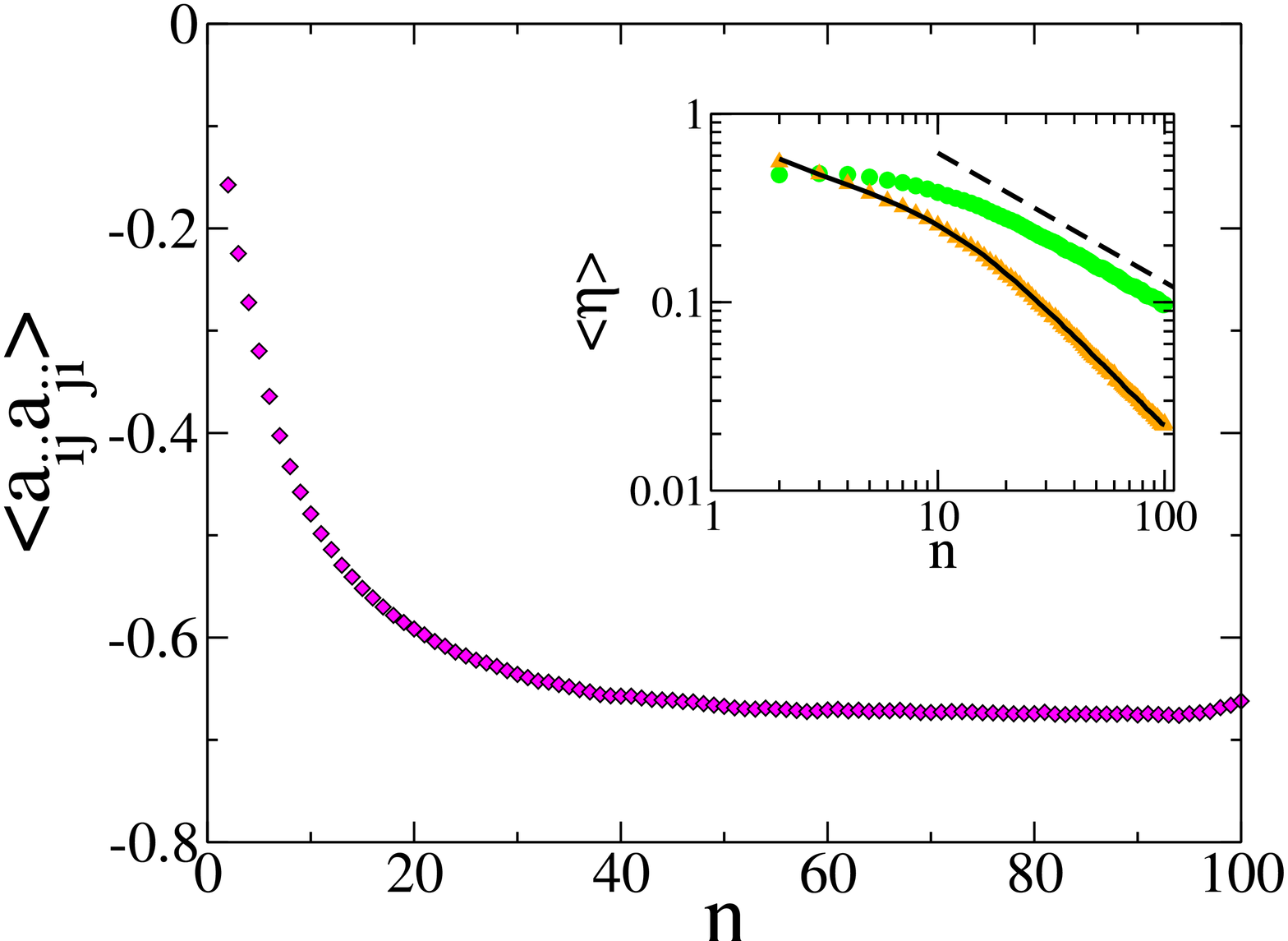}
\caption{\label{aijn}  Correlation function $\langle a_{ij}a_{ji} \rangle$ between a pair of nodes $(i,j)$ with a double link as a function of the network size, for $b=2$ and $q=3$. The average was calculated over all pair of sites with double link in networks with the same size. The inset shows a comparison between the fraction of double links in the present network $\langle \eta \rangle$ (green circles) and a random network of the same size and connectivity $C(n)$: $\langle \eta \rangle_{ran}=C/(2-C)$ (continuous line). Yellow  triangles correspond to a numerical calculation of $\langle \eta \rangle_{ran}$. The dashed line corresponds to a power law $n^{-0.68}$.}
\end{center}
\end{figure}

In Fig.~\ref{aijn}, we show $\langle a_{ij}a_{ji} \rangle$ as a function of the network size $n$. We see that this correlation is negative for any value of $n$ and saturates into a value $\langle a_{ij}a_{ji} \rangle \approx -0.65$ for large values of $n$. In the inset of Fig.~\ref{aijn}, we compare the average fraction of double links  $\langle \eta \rangle$ with the corresponding quantity for a completely random network with the same connectivity $C(n)$, that is, a network where all edges are independently distributed with probability $P(a_{ij}\neq 0)=C(n)$. Then, the probability of having a link between an arbitrary pair of sites is $p_d =  C(n)(2-C(n))$
 and the average degree per node $\langle k \rangle = p_d n$. Hence

 \begin{align}\label{etarand}
 \langle \eta \rangle_{ran}= \frac{C^2(n)\,n}{\langle k \rangle} = \frac{C(n)}{2-C(n)}
 \end{align}

\noindent Then for large values of $n$, we have $\langle \eta \rangle_{ran}\sim C(n) \sim n^{-(1+\epsilon)}$. From the inset of Fig.~\ref{aijn}, we see that $\langle \eta \rangle \sim n^{-0.68}$ when $n \gg 1$ in the present case.  The fraction of double links  is considerable larger than in a random network. The two results of Fig.~\ref{aijn} together show that the present networks have indeed  a significantly large number of anticorrelated pair interactions.

\begin{figure}
\begin{center}
\includegraphics[scale=0.27,angle=-90]{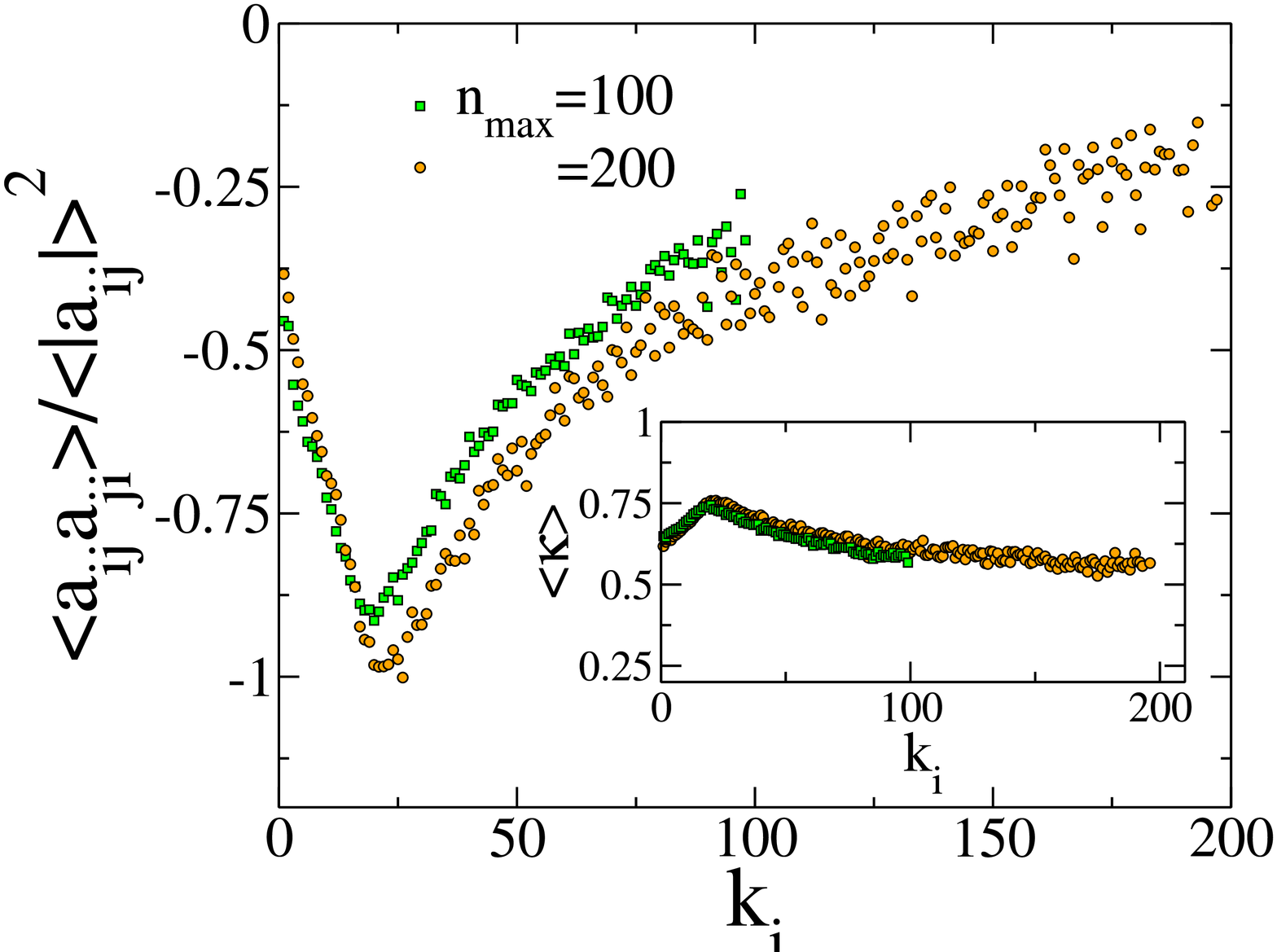}
\caption{\label{aijk}  Correlation function $\langle a_{ij}a_{ji} \rangle/\langle|a_{ij}| \rangle^2$ between the matrix elements linking a node $i$ and its neighbors $j$ as a function of its degree $k_i$, for $b=2$, $q=3$ and different values of $n_{max}$. The inset shows the average fraction of anticorrelated links $\langle \kappa \rangle$  as a function of $k_i$.}
\end{center}
\end{figure}

Next, we calculated
the  correlation $\langle a_{ij}a_{ji} \rangle/\langle|a_{ij}| \rangle^2$  between the matrix elements, linking a node $i$ and its neighbors $j$, as a function of its degree   $k_i$, where the average is taken only on the double links.
 From Fig.~\ref{aijk}, we see that the absolute value of the correlation presents a maximum around $k_i= 25$ and tends to zero as the degree increases.  The inset of Fig.~\ref{aijk} shows that the average fraction of anticorrelated links $\langle \kappa \rangle$ (i.e., $\#$ anticorrelated links/total $\#$ double links) tends to $1/2$ as the degree increases. We can conclude from these results that the interactions strengths between the hubs and their neighbors are almost uncorrelated. This suggests that the influence of hubs in the stabilization of the dynamics is mainly associated to their topological role (e.g., reduction of the average length $L$) rather than to the nature of their associated interactions.

\section{Discussion}
\label{conclu}

 The recent  advances  in the research on networks theory in biological systems have called for a deeper understanding about the relationship between network structure and function, based on evolutionary grounds \cite{PrPrPh2005}. In this work,
 we have shown that a key factor to explain the emergency of many of the complex topological features commonly observed in biological networks could be just the stability of the underlying dynamics. Stability can then be considered as an effective fitness acting in all biological situations. The results presented in  Fig.~\ref{Cndata} for the connectivity of real biological networks at different network size scales support this conclusion.  In addition, the present approach (although based on a very simple model) allows to draw some conclusions about the interplay between network structure and function that could be of general applicability. The present results suggest that hubs play mainly a topological role of linking modules (disassortativity, low clustering, uncorrelated links), while low connected nodes inside modules enhance stability through the presence of many anticorrelated interactions.

The stabilizing effects of some of the topological  and functional network features here analyzed have been previously addressed separately (small world \cite{Si2005}, dissasortative mixing \cite{BrSi2005,SiSi2006}, anticorrelated interactions \cite{AlPa2008}). However, the present analysis suggests that the simultaneous observance of all of them is highly unlikely to be a result of a purely random process. Such delicate balance of specific topological and functional features would only be attainable through a slow, evolutionary stability selection process.

In particular, the above scenario agrees very well with the observed structures in cellular networks. For instance, the scaling behavior of $Cc(k)$, displayed in Fig.~\ref{Cck}, has been observed in metabolic \cite{RaSoMoOlBa} and protein \cite{YoOlBa2004,CoFlMaVe2005} networks. Disassortative mixing by degree is another ubiquous property of those systems and indeed a very similar behavior to that shown in Fig.~\ref{Knn} has been observed in certain protein-protein interaction networks \cite{CoFlMaVe2005}. Also, the available data for the degree distribution in all those cases are consistent with a power law behavior with an exponent $\gamma$ between $2$ and $2.5$ \cite{AlBa2002}. The agreement with the whole set of properties predicted by the model suggests that stability could be a key evolutionary factor in the development of cellular networks.

The situation is a bit different in the case of ecological networks, where the predictions of the model do not completely agree with the observations, specially those related to food webs. On the one hand,  food webs usually display also disassortative mixing by degree, modularity and relatively low small worldness \cite{Du2006} (rather low values of clustering, compared with other biological networks), in agreement with the present predictions. Regarding  the scaling behavior of $C(n)$ \cite{Note3}, this is the topic of an old debate in ecology (see Dunne's review in Ref. \cite{Du2009} for a summary of the debate). While in general it is expected a power law behavior, the value of the exponent (and the associated interpretations) is controversial, due to the large dispersion of the available data, the rather small range of network sizes available and, in some cases, the low resolution of the data \cite{Note4}. The consistency of the scaling shown in Fig.~\ref{Cndata} for a broad range of size scales suggests that the ecological debate should be reconsidered in a broader context of  evolutionary growth under stability constraints.

On the other hand, the degree distribution of food webs is not always a power law, but it frequently exhibits an exponential cutoff at some maximum characteristic degree $k_{max}$ \cite{MoPiSo2006,DuWiMa2002}. Such variance between food webs and other biological networks is probably related to the way ecosystems assemble and evolve compared with other systems.
While the hypothesis of the present model are general enough to apply in principle to any biological system, that difference suggests that stability is not enough to explain the observed structures in food webs, but further constraints should be included to account for them. For instance, at least two different (although closely related) constraints are known that can generate a cutoff in the degree distribution: aging  and limited capacity of the nodes \cite{AmScBaSt2000}. In the former, nodes can become inactive with some probability through time (in the sense that they stop interacting with new agents), while in the latter they systematically pay a ``cost'' every time a new link is established with them, so that they become inactive when some maximum degree is reached. One can easily imagine different situations in which  mechanisms of that type become important in the evolution of ecological webs, either by limitations in the available resources or by dynamical changes in the diet of species due to external perturbations (for instance, there are many factors that constrain a predator's diet; see Ref. \cite{MoPiSo2006} and references therein for a related discussion). Mechanisms of these kind can be easily incorporated into the model, serving as a base for the description of more complex behaviors in particular systems like food webs.

Finally, it would be interesting to analyze the relationship between dynamical stability  in evolving complex networks and synchronization, a topic about which closely related results have been recently published \cite{NiMo2010}.

\begin{acknowledgements}
 This work was
supported by CONICET, Universidad Nacional de
C\'ordoba, and FONCyT grant PICT-2005 33305 (Argentina). We thank useful discussions with P. Gleiser and D. R. Chialvo. We acknowledge  useful comments and criticisms of the referee.
\end{acknowledgements}


\begin{thebibliography}{9000}
\bibitem{Note-food-webs} In particular cases, like food-webs, there could be other coupled process almost probably as important as the two considered here, such as the dynamical change in previously existing links weights. For instance, if the abundace of one species is reduced, this could trigger the adaptation of predators to feed on other existing species, thus creating new links not directly related to the incoming species. The present scheme should be considered as a minimal model about stability as an evolutionary constraint.
\end{thebibliography}
\end{document}